\documentclass[a4paper,12pt,oneside]{article}
\usepackage[utf8]{inputenc} 
\usepackage[T2A]{fontenc}
\usepackage[english]{babel}

\usepackage{amsmath}
\usepackage{amssymb}
\usepackage{latexsym}
\usepackage {xcolor}
\usepackage[table]{colortbl}
\usepackage{graphicx}
\usepackage{epsfig}
\usepackage{caption}
\usepackage{wasysym}
\usepackage[section,above,below]{placeins}
\usepackage{cite}
\usepackage[colorlinks,citecolor=blue,linkcolor=red]{hyperref}

\title{Towards High-Power Microwaves}
\author{S.Anishchenko, V.Baryshevsky, A.Gurinovich,\\ 
	E.Gurnevich, P.Molchanov, A.Rouba.\\ 	\vspace{0.1cm} \\
\textit{Institute for Nuclear Problems of Belarusian State University \&} \\\textit{Electrophysical laboratory, Minsk, Belarus}}

\begin{document}
\maketitle
\begin{abstract}
%Review and comparison of HPM sources operating without magnetic field, used to guide an electron beam, and capable to produce high-power microwave pulses with duration about 100 ns. The proposed analysis summarizes multi-year research carried with three types of
%HPM sources, namely: axial vircator, split-cavity oscillator and virtual cathode oscillator in reflex triode geometry.
%These options were simulated for electron beam energy $\sim$400 keV and demonstrated capability to provide high power microwave pulses with the required pulse duration. Designed sources are experimentally tested, their advantages and weaknesses are discussed considering high
%output power, long pulse duration and good operation stability.

%%todo from reviewer: --------- >>

In this paper, we review and compare HPM sources operating without a magnetic field to guide the electron beam that are capable of producing high-power microwave (HPM) pulses with a duration of about 100 ns. The proposed analysis summarizes multi-year research carried with three types of HPM sources: a split-cavity oscillator (SCO); an axial vircator;  and a virtual cathode oscillator in Reflex Triode geometry. These options were simulated for electron beam energy $\sim$400~keV and for pulsers with demonstrated capability to provide high power microwave pulses with the required pulse duration. Designed sources were experimentally tested, and their advantages and weaknesses are discussed with respect to high output power, long pulse duration, and good operating stability.

%%todo end from reviewer: << ---------

\end{abstract}

\section{Introduction}

The aim of the presented analysis is to compare HPM sources of different types and substantiate the choice 
of a source that will provide high output power, long pulse duration and good operating stability. 
The technical requirements for the HPM source are operation without a magnetic field and compact design. Single frequency operation during the 100~ns pulse is desired.
%

%%todo change the sequence in accordance with the sequence in main text
Three types of HPM sources which comply with the above features are analyzed and compared, namely: 
a split-cavity oscillator (SCO), an axial vircator  and a virtual cathode oscillator in Reflex Triode geometry (referred hereinafter as reflex triode). 
%Operation without guiding magnetic field implies short  length of beam-wave interaction 
%area that is beneficial for source compactness, but simultaneously purports quite low efficiency. 

Split-cavity oscillator (SCO), as proposed in \cite{Marder1992}, is a compact device, 
whose self-excited oscillating electromagnetic field converts a steady electron beam into 
one with highly modulated density. Short travel length eases overcoming both 
space-charge and pinching limitations, thus, enabling  high currents and quite high input power. 

Devices based on the virtual cathode principle operate at a current density 
that exceeds the space charge limit \cite{Miller1982,Granatstein1987,Benford2007,Didenko1979,Sullivan1983}. 
High input power somewhat compensates low intrinsic efficiency, which nevertheless could be 
increased by inserting resonant elements into the beam-wave interaction area.
An HPM source with a virtual cathode is usually composed of a few elements: vacuum diode 
with a cathode emitting electrons and mesh anode letting the electron beam to enter the 
resonant cavity, which design could be elaborate though.

Differences in operation mechanism, electron beam dynamics, source impedance and radiation mode 
for all three types of sources make comparison challenging. 
We have chosen the following parameters for comparison: 
output radiation power, 
duration of radiated pulse, 
operation stability (reproducibility of results).

Each of three sources we tested was designed to operate at electron beam energy $\sim$300--500~keV.
%
%%TODO translate
%Типичные значения токов в экспериментах с SCO составляли от 3 до 8~kA,  
%для систем с виртуальным катодом токи были существенно выше: $\sim 13-20$~kA.
The typical current values for SCO ranged from 3 to 8~kA, for virtual cathode oscillators they were expectedly higher:  13-20~kA.

Barrel-type
housing rendered volume available for evacuation and placing the structure (resonator), inside which 
the electron beam interacts with electromagnetic field.
HV was applied to the core conductor from the high-voltage pulse source; vacuum feed-through 
isolator bounded the evacuated volume from one face, another face was covered by the output window.
%todo make the Figure from drawing with HV isolator and housing
HV pulse polarity, either negative for axial vircator and SCO or positive for reflex triode, was 
determined by the power supply.
The applied voltage  and the beam current were measured to evaluate the input power.  
Source impedance was varied by changing the cathode-anode gap and the diameter of cathode. 
Cathodes of different types were tested, resonant elements in composition of each source were used 
for fine tuning radiation output.

Radiated electric field strength was measured in the direction of its maximal value at some distance 
from the output window and reduced to 1~meter distance. 
Output power was evaluated considering HPM source directivity and was used for efficiency calculation.
%%TODO translate
Output power values in this paper are averaged over several oscillation periods. 
%
%значения мощности излучения в данной статье приводятся как мощность усредненная по нескольким периодам генерации).
Radiated time-domain waveform was analyzed to reveal its spectral content and power distribution 
within the pulse duration interval.

This paper is focused on summarizing and comparing the results obtained 
experimentally.  %The paper is organized as follows. 
Each subsequent  section \ref{sec:SCO} - \ref{sec:reflex} presents the experimentally obtained results 
and discusses the observed advantages and weaknesses of each source. 
Conclusive section discusses comparison results.

\section{SCO}
\label{sec:SCO}

The split-cavity oscillator proposed in ~\cite{Marder1992} was operating at 100-150~kV to radiate at $\sim$1~GHz.
A high-quality pillbox cavity with conducting screen walls, through
which an electron beam could pass, 
was partitioned (split) by a screen leaving a gap between it and
	the outer cavity wall.
The self-excited  electromagnetic
field oscillating at the frequency of fundamental mode of the split cavity enabled converting a steady electron beam into highly density modulated one.
Conducting screen walls and split meshes also served for beam guiding: radial electric field component was reduced, thus, reducing Coulomb repulsion in the beam without the need for an applied magnetic field.
The extractor collected the exit beam
on an inverse diode in a wave matching area and launched an electromagnetic wave
into a coaxial transmission line and output horn.

A disadvantage of the above SCO was the strong compression of electron beam ~\cite{Lemke1994} exiting the pillbox by its own magnetic field that caused plasma emission from the extractor face due to electron bombardment and, thus, radiation failure.

Experimental studies \cite{Fan2004,He2004} demonstrated possibility of SCO operation at higher frequencies at 400 -- 500~kV producing power 200--400~MW. 
Operation of SCO at high beam energies was considered in \cite{Bar14}.
SCO cavity split to unequal parts was studied in \cite{He2004,Moroz2018,Moroz2020}.

\begin{figure}
	\leavevmode
	\centering
	\resizebox{160mm}{!}{\includegraphics{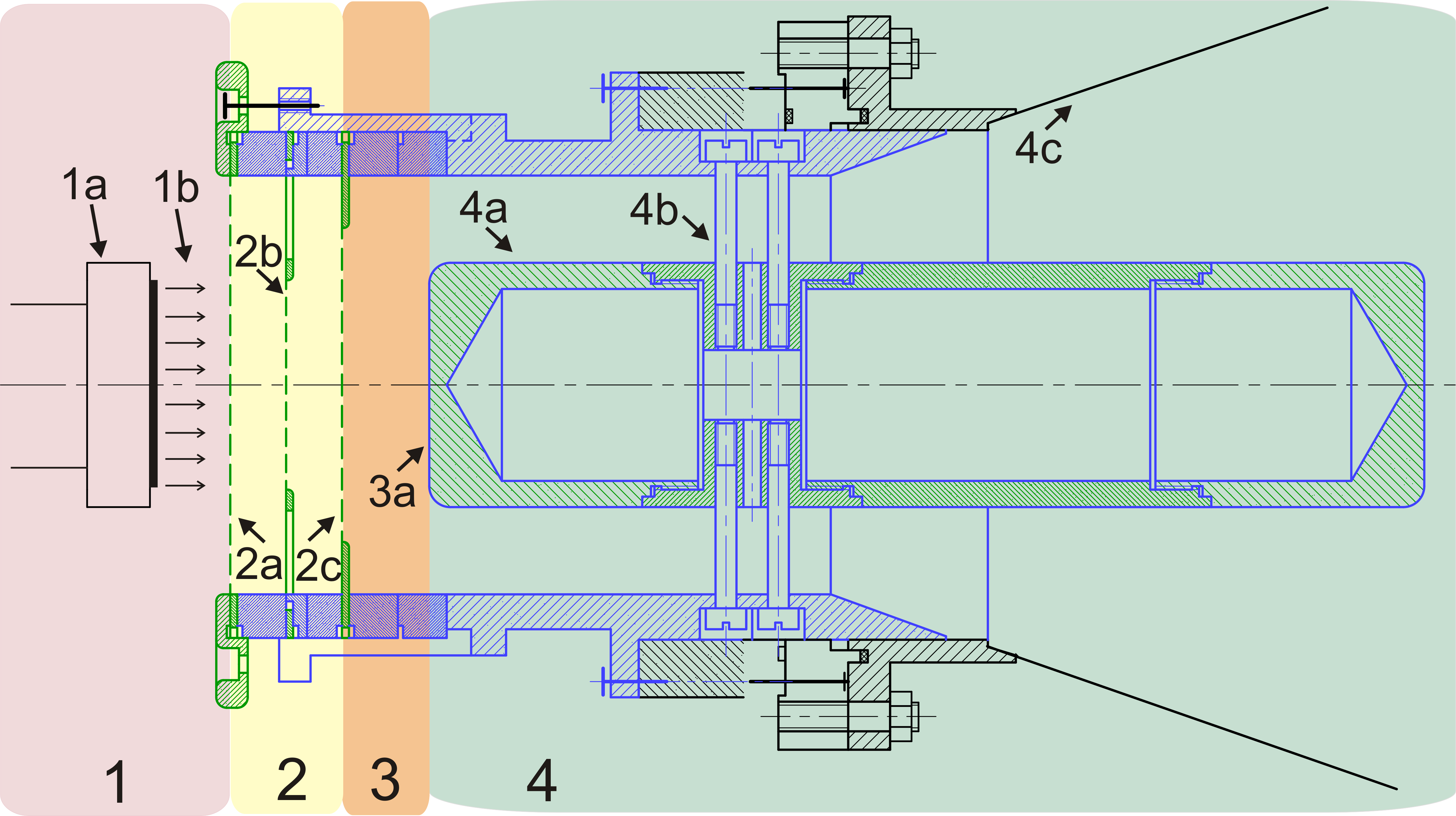}}
	\captionof{figure}{SCO design option~\#1: 
		\newline 1 -- vacuum diode; 1a -- graphite cathode $\diameter$~50~mm, 1b -- cathode-anode gap 14~mm; 
		\newline 2 -- pillbox~\#1 with length~28~mm and inner $\diameter$ 120~mm: 2a -- anode mesh (mesh \#1), 2b -- split mesh with $\diameter$90~mm, 
		2c -- mesh \#2 with $\diameter$90~mm with support ring;
		\newline 3 -- pillbox \#2 with length~20~mm and inner $\diameter$ 120~mm:
		3a -- extractor face serving to absorb e-beam; 
		\newline 4 -- wave matching area:
		4a -- extractor with length~285~mm, $\diameter$70~mm, 
		4b -- extractor supports, 
		4c -- horn with length~550~mm, $\diameter$480~mm }
	\label{fig:scomarder}
\end{figure}

\begin{figure}[htp]
	\centering
	\includegraphics[angle=0,width=0.5\linewidth]{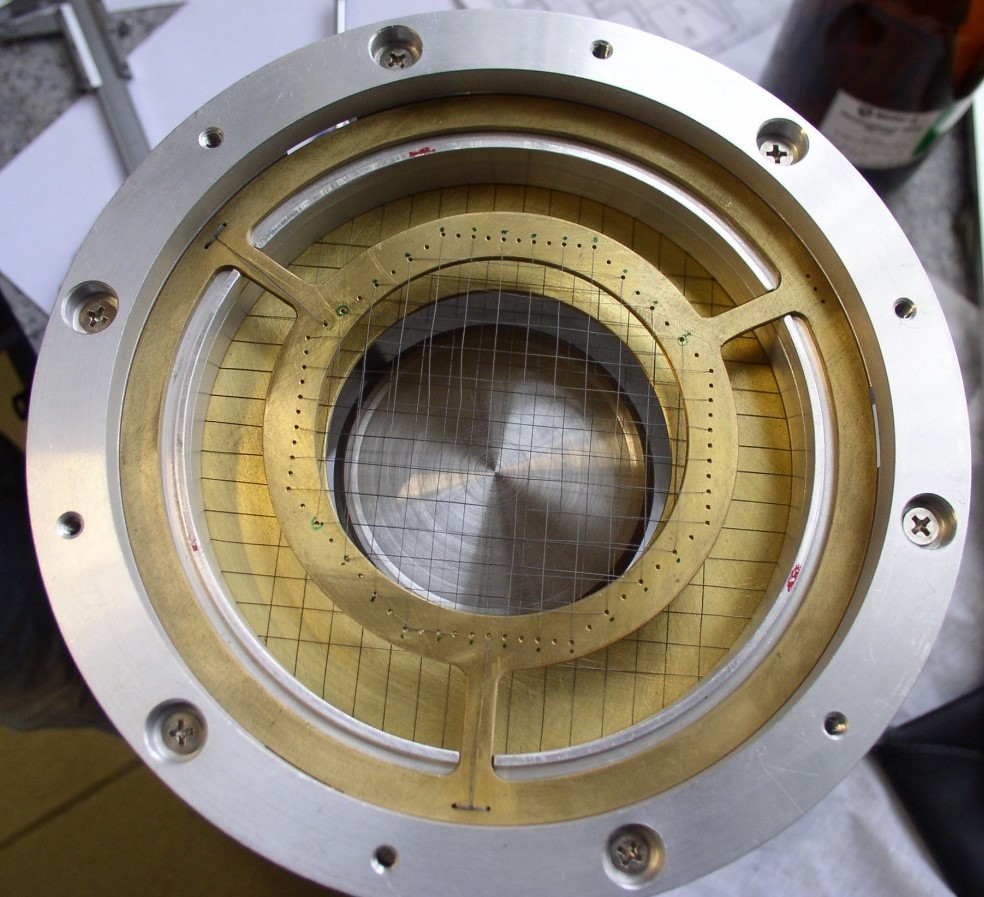}
	\caption{Disassembled SCO design option~\#1, split mesh and metal extractor can be seen (see Fig.~\ref{fig:scomarder})}
	\label{fig:mesh2}
\end{figure}

\begin{figure}
	\leavevmode
	\centering
	\resizebox{160mm}{!}{\includegraphics{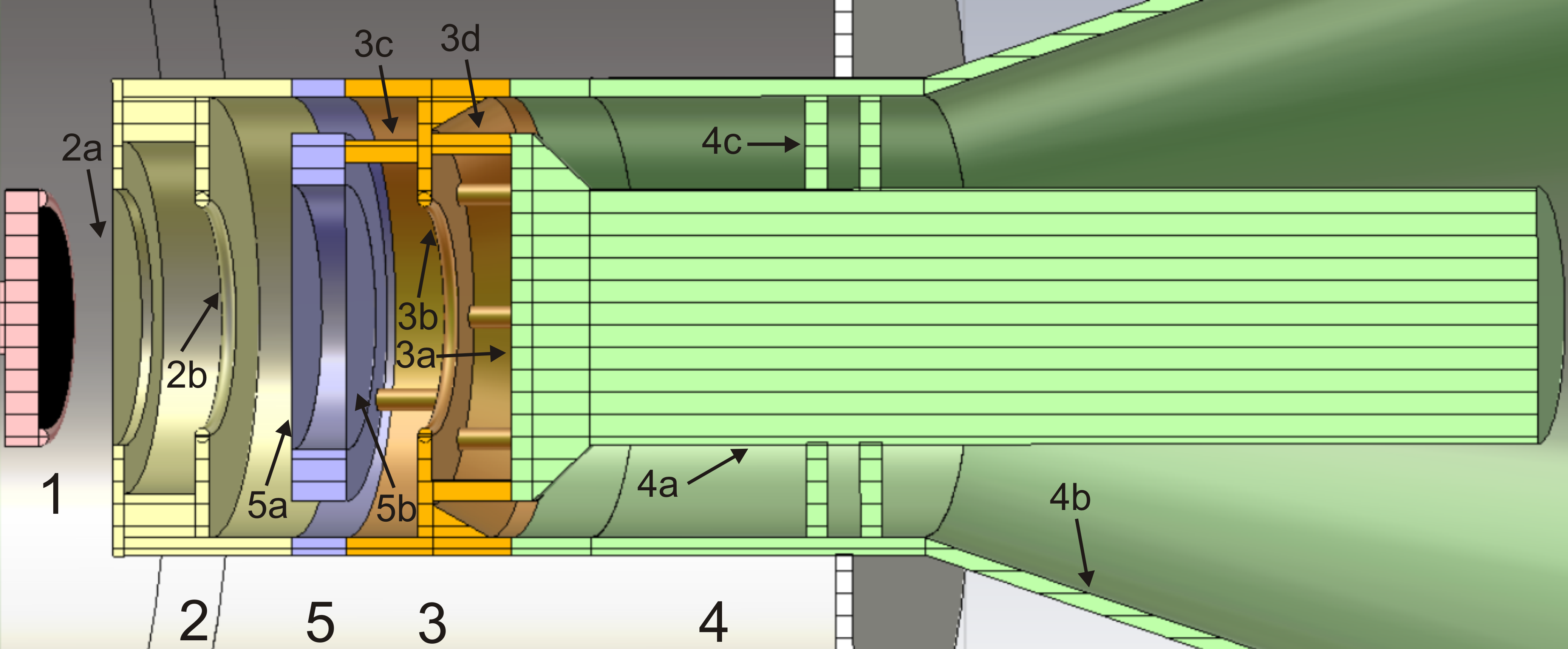}}
	\captionof{figure}{
		SCO design option~\#2 (with split diaphragms): 
		\newline 1 -- vacuum diode with carbon-multicapillary cathode $\diameter$~60~mm and cathode-anode gap 16-20~mm; 
		\newline 2 -- pillbox~\#1 with length~50~mm and inner $\diameter$ 120~mm: 2a -- anode mesh (mesh \#1), 2b -- split diaphragm \#1 with inner $\diameter$60~mm;
		\newline 3 -- pillbox \#2 with length~44~mm and inner $\diameter$ 120~mm: 3a -- extractor face serving to absorb e-beam, 3b -- split diaphragm \#2 with inner $\diameter$60~mm, 3c -- drift chamber supports, 3d -- metal extractor supports; 
		\newline 4 -- wave matching area:
		4a -- extractor with length~285~mm, input/output $\diameter$100/70~mm,		%
		4b -- horn with length~550~mm, $\diameter$480~mm,
		4c -- dielectric extractor supports; 
		\newline 5 -- drift chamber with length~15~mm, outer/inner $\diameter$100/72~mm:	
		5a -- mesh~\#2, 5b -- mesh~\#3}
	\label{fig:sco}
\end{figure}

We designed  and tested two split-cavity oscillators to operate at 300--400~kV and $\sim$~3~GHz using the approaches \cite{Marder1992,Lemke1994,Fan2004,He2004,Bar14}. 
Similar \cite{Marder1992} the split mesh was used in SCO design option~\#1   (see Fig.\ref{fig:scomarder},\ref{fig:mesh2}), while in the design option~\#2 we replaced mesh with a split diaphragm (see Fig.\ref{fig:sco}). %
%

%\begin{minipage}{0.5\textwidth} ---------------

%\begin{minipage}{0.5\textwidth}

\begin{figure}[ht]
	\leavevmode
	\centering
	\resizebox{40mm}{!}{\includegraphics{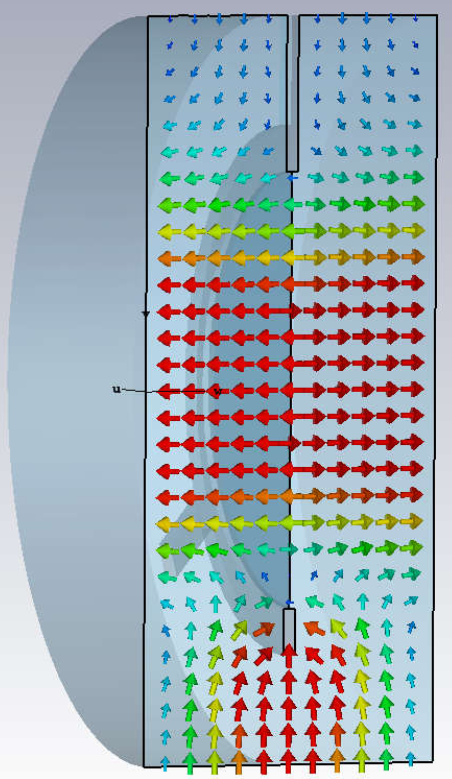}}
	\resizebox{40mm}{!}{\includegraphics{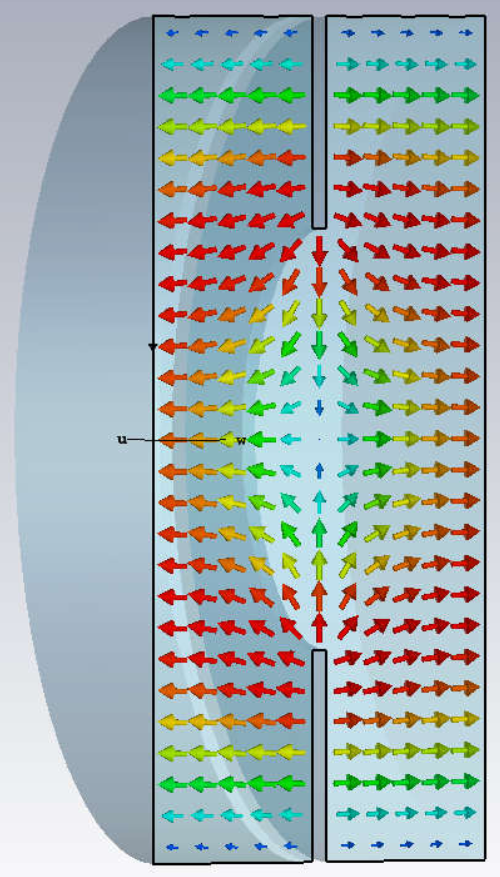}}
	\caption{RF electrical field of the fundamental mode in SCO cavity design option \#1 (left) and design option \#2 (right)}
	\label{fig:E_SCO_TM}
\end{figure}

In SCO design option~\#1  pillbox was used for electron beam modulation and wave matching area between SCO and extractor served for beam energy conversion to radiation.

SCO design option~\#2  was composed of two pillboxes with a drift chamber between them: the first pillbox in this case modulated the beam, the second one served for beam energy conversion to radiation.
% similar \cite{Fan2004,He2004} 
%
In contrast to  \cite{Fan2004,He2004} we used split diaphragms instead of split meshes. Such replacement enabled to reduce multiple scattering of beam electrons by the meshes and to eliminate all the associated effects. 
Additionally, electron beam compression was reduced. 

Coupling of two pillboxes was enabled by design of drift chamber 
placed with a gap between it and the outer cavity wall that enabled  field oscillation in a joint volume without special SCO cavities tuning.
	
For SCO design option~\#1 we tested two variants of extractors: that purely metal and covered with graphite.  
Applying graphite cover to an extractor is beneficial for output power because of lower albedo for graphite as compared to metal \cite{Bogd2004}. Therefore for SCO design option~\#2 we used only  extractor covered with graphite.

Similar \cite{Fan2004,He2004} duration of radiation pulse (approximately 10~ns) in our experiments was much shorter than current pulse width, which was about 100~ns.  
We explained this by plasma formation inside the cavity and wave matching area;  use of duralumin for manufacturing the SCO parts exacerbated the situation  because of oxide film on all the surfaces. Multiple scattering and secondary electron emission from all the parts bombarded by the electrons also contributed. 

Input and output parameters for both SCO design options are gathered in Table \ref{tab:SCO_e}.

\begin{figure}[ht]
	\centering
	\leavevmode
	\resizebox{150mm}{!}{\includegraphics{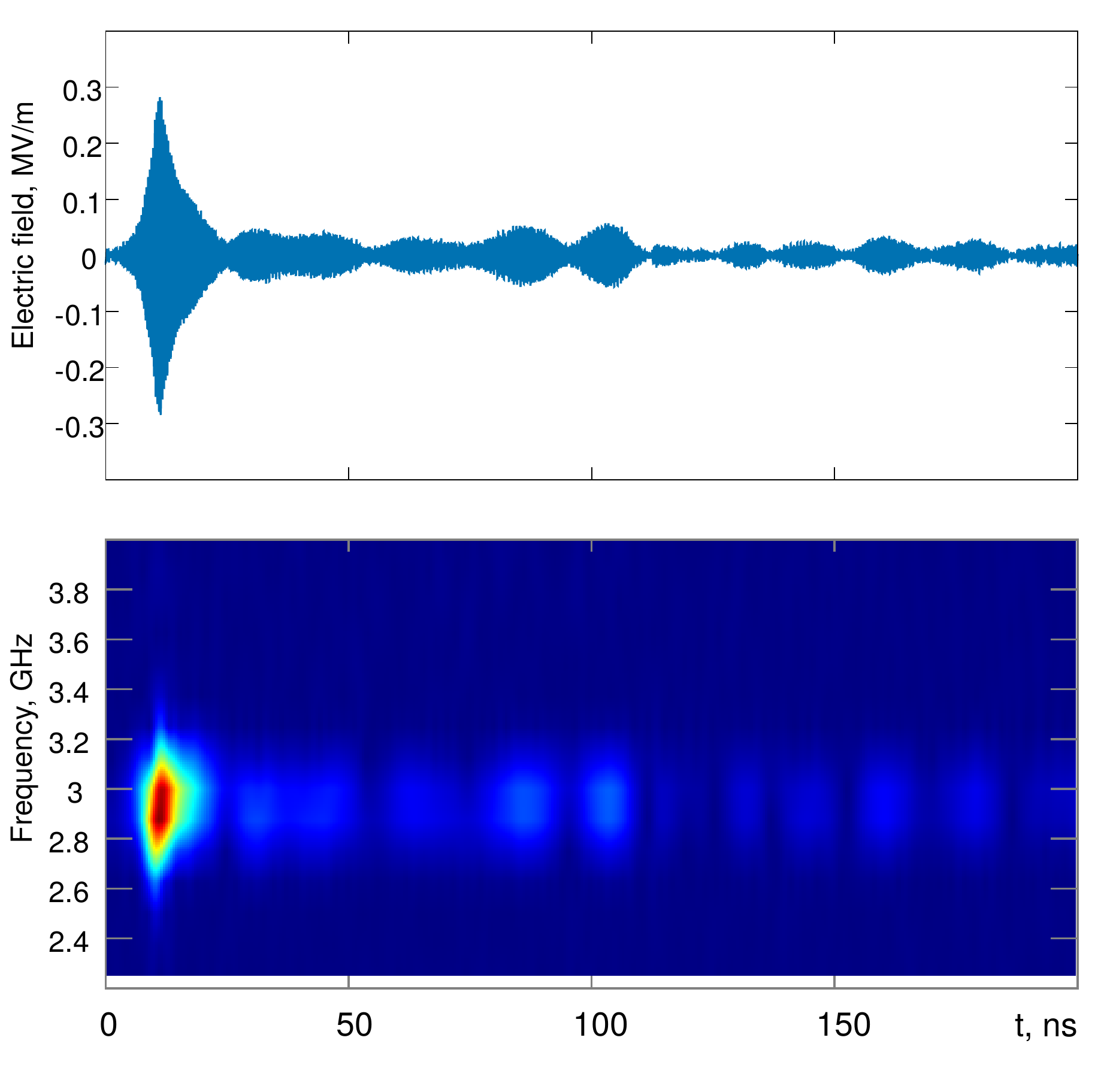}}\\
	\caption{Detected microwave signal with the highest amplitude and its spectrogram for SCO design option \#2}
	\label{fig:esco}
\end{figure}

\begin{table}[htp]
	\caption{Input and output parameters of HPM sources on the base of SCO.}
	\label{tab:SCO_e}
	\begin{tabular}{|>{\centering\arraybackslash}m{0.5cm}|>{\centering\arraybackslash}m{6.5cm}|>{\centering\arraybackslash}m{3.5cm}|>{\centering\arraybackslash}m{3.5cm}|}
		\hline
		\textbf{No.} & \textbf{Parameter}		&\textbf{SCO design option~\#1}	&\textbf{SCO design option~\#2}
		\\ \hline
%		1		& \multicolumn{3}{c|}{HV Power supply}
%		\\ \hline
%		& Type 	&\multicolumn{2}{c|}{Marx generator}  
%		\\		\hline
	1	& Voltage at start of HPM generation 		& 300 kV	&  	350 kV				
		\\ \hline
	2	& Current at start of HPM generation 	&	8 kA	&  	4 kA				
		\\ \hline
%		2		& \multicolumn{3}{c|}{HPM Source}
%		\\ \hline
	3	& HPM frequency 	&\multicolumn{2}{c|}{$\sim$ 3 GHz}  
		\\		\hline 
	4	& E-field @ 1 meter   	&  	300 kV/m @1m	&  	250 kV/m @1m				
		\\ \hline
	5	& Raduated pulse FWHM 		&  	$\sim$10 ns	&  	$\sim$5 ns				
		\\ \hline
	6	& Power density	&  	125 MW/m$^2$	&  	85 MW/m$^2$			
		\\ \hline
	7	& Power		&  	\multicolumn{2}{c|}{$\sim$ 30 MW} 		
		\\ \hline
	8	& Efficiency	&  	1.3 \% & 2.1 \%    			
		\\ \hline
	\end{tabular}
\end{table}

%\begin{figure}[ht]
%	\leavevmode
%	\centering
%	\resizebox{80mm}{!}{\includegraphics{img/scoSignal.eps}}\\
%	\caption{SCO.}
%	\label{fig:scoSignal}
%\end{figure}

%\begin{figure}[ht]
%	\leavevmode
%	\centering
%	\resizebox{80mm}{!}{\includegraphics{img/scoSpectrum.eps}}\\
%	\caption{SCO.}
%	\label{fig:scoSpectrum}
%\end{figure}

%todo remove subsection
%\subsection{Cathode-anode gap}

\begin{figure}[th]
	\centering
	\resizebox{140mm}{!}{\includegraphics{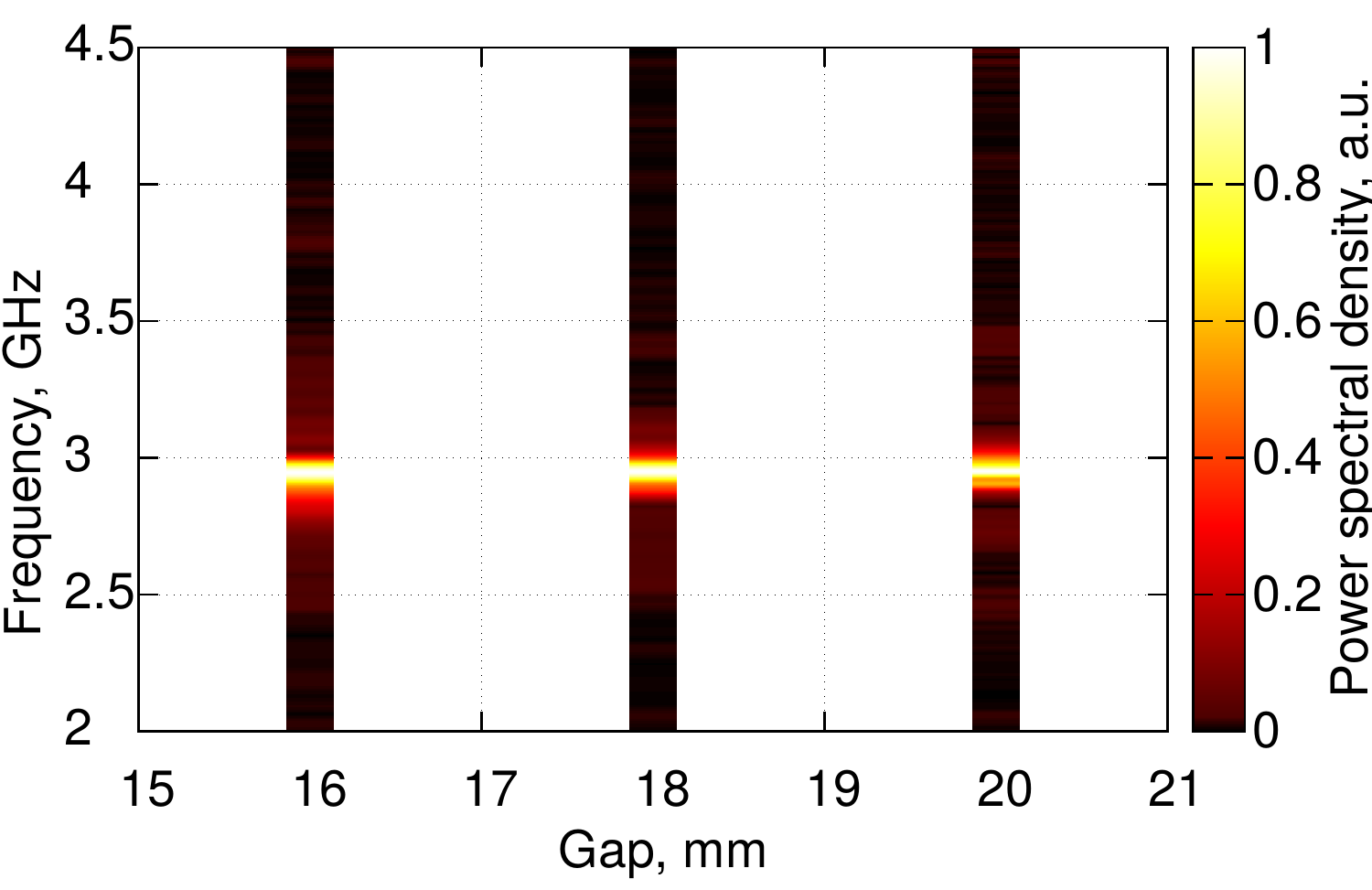}}\\
	\caption{Comparison of spectra for SCO with split diaphragms obtained for different cathode-anode gap values}\label{fig:specs_sco}
\end{figure}

RF electrical field of the fundamental mode in SCO cavity  
determines SCO operation frequency (comparison of field structure for SCO 
with the split mesh and that for the cavity with the split diaphragm is given in Fig.~\ref{fig:E_SCO_TM}). That is why change of the gap between the anode and the cathode does not influence the radiation frequency (see Fig. \ref{fig:specs_sco}). Gap change can influence radiation only via impedance and beam current change.

%%TODO translate

Experiments with SCO demonstrated stable single-frequency operation. 
Detected output power was as high as 30~MW and efficiency was $\sim$2\%.
However, 
the duration of radiation pulses were much shorter than the width of the current pulse (see Table~\ref{tab:SCO_e}) despite the fact that start of HPM generation occured at growing voltage and current in vacuum diode (well before the maxima of voltage and current).  
%Длительность микроволновых импульсов $\sim 5$~нс значительно меньше длительности 
%токового импульса ($\sim200$~нс), что свидетельствует о наличии преждевременных пробоев в системе. 

%Несмотря на неудовлетворительные результаты эксперимента и понимание их причины, дальнейшее исследование по данному направлению было решено прекратить. 
%%todo translate
Pulse shortening was explained by enhancement of electric field within small gaps inside the SCO structure and plasma therein. Power growth could also be limited by breakdown, similar reasons limiting MILO operation in~\cite{Calico,Haworth1998}.
%Анализ показал, что слишком малые зазоры между элементами электродинамической структуры, сильные электрические поля не позволят существенно удлиннить импульс излучения. По всей видимости, схожие причины ограничивают работу MILO~\cite{Calico,Haworth1998}.

%%todo translate
All the measures reducing multiple scattering, secondary electron emission and electric field strength inside SCO structure would be beneficial for both output power and pulse duration. SCO designed for operation at lower  frequency has some advantages, since the dimensions of SCO structure and all the gaps are expected to be larger.  
%Возможными решениямими для увеличению длительности и мощности излучения являются уменьшение вторичной электронной эмиссии внутри генератора, снижение напряженности электрического поля в SCO, снижение  рабочей частоты и, как следствие, увеличения расстояния между элементами SCO.  
%%---------- %TODO translate

{Simulated and experimental results showed that the stability of the observed results was high, but the experimental results showed that the radiated power and pulse duration were low when compared to the simulation. Analysis of the results enabled us to assume that pulse shortening was being caused by secondary electron emission inside the SCO and by the expanding plasma cloud, which is formed in the vicinity of the extractor as a result of its being bombarded by electrons causing the beam to be compressed by its own magnetic field.}

\section{Axial vircator}
\label{sec:axial}

Different vircator designs with resonant cavities were proposed \cite{Jiang2004,Kitsanov2002,Didenko1996,Champeaux2013}.
The studies of axial vircators with multicavity resonators proved the possibility of a resonant increase in the efficiency \cite{Zhenxiang2003,Ting2002,ZhiQiang2008}.

\begin{figure}[ht]
	\leavevmode
	\centering
	\resizebox{150mm}{!}{\includegraphics{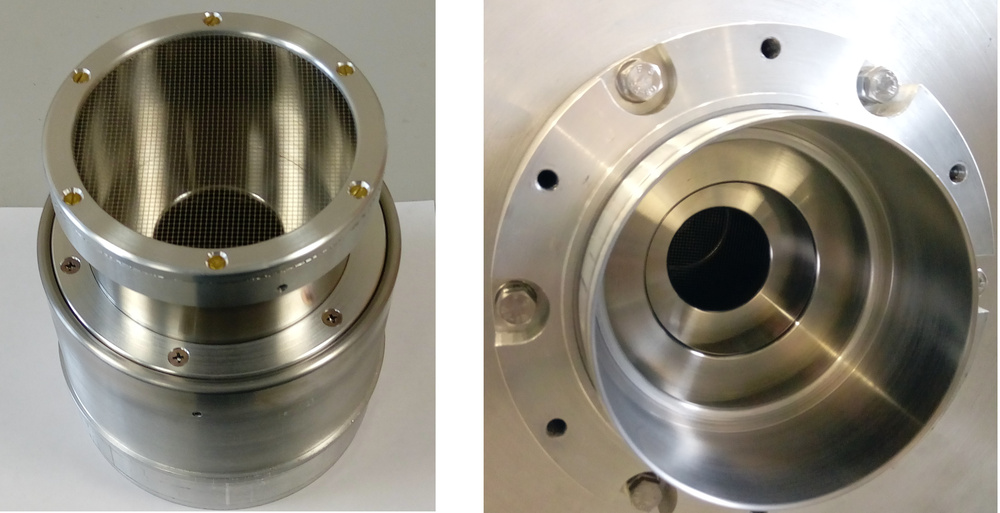}
		\hfil
\includegraphics[height=18.5 cm]{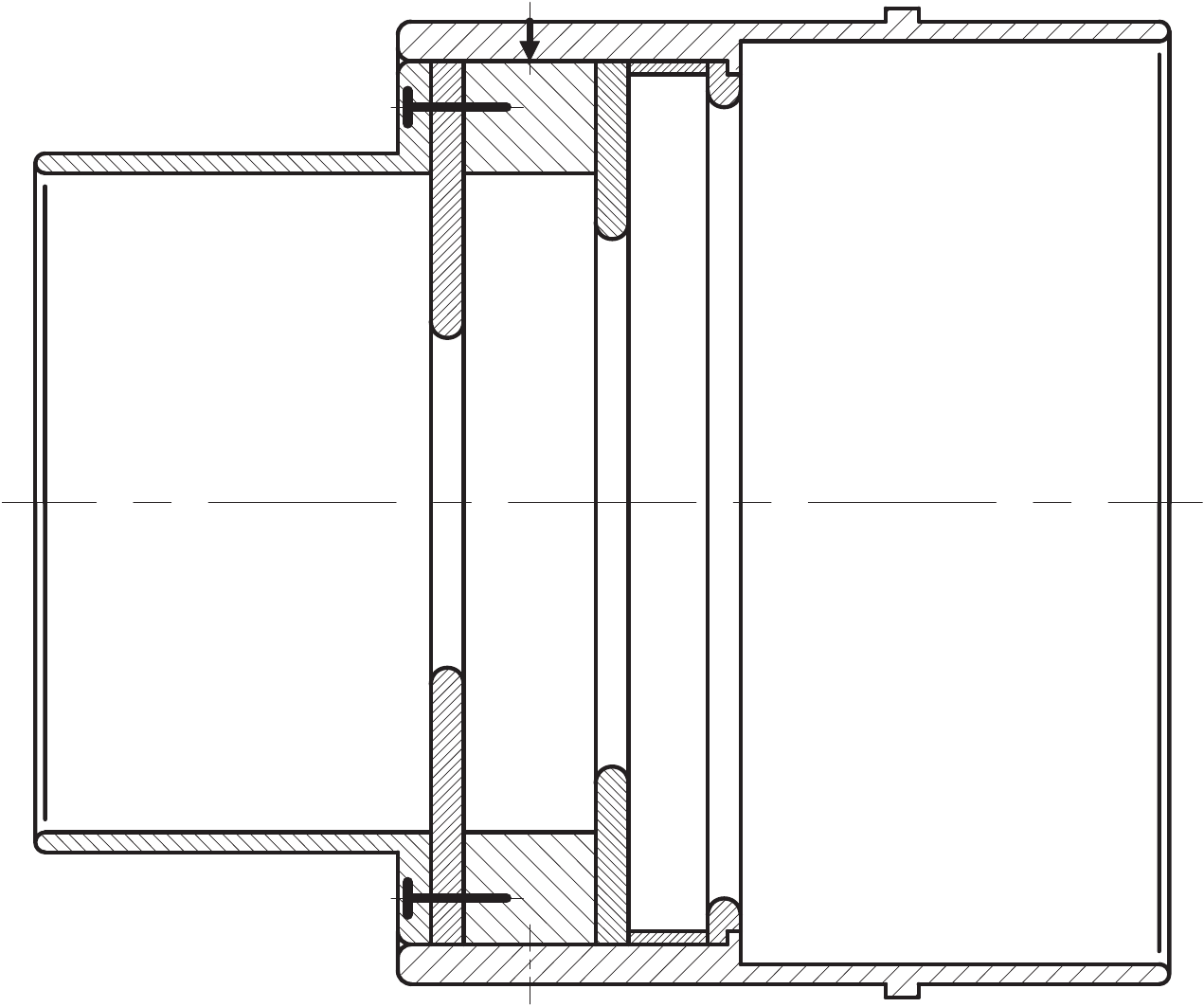}	
}\\
	\caption{Three cavity resonator of axial vircator~\cite{Baryshevsky2015}: outside view with mounted anode mesh (left), inside view (mesh is unmounted, middle), drawing (left)}
	\label{fig:axialresonator}
\end{figure}

Using \cite{ZhiQiang2008}  as a starting point for the development, we designed several resonators for the 
three-cavity axial vircator \cite{Baryshevsky2015, Molchanov2014} (see one of them in Fig.~\ref{fig:axialresonator}).
In simulation~\cite{Molchanov2014,Gurnevich2015} the axial vircator  designed for operation in the 
frequency range from 3 to 4 GHz at 450 keV
demonstrated efficiency  greater than 5\% and the power reached 350~MW at 3.2~GHz
(mind that both simulated and observed values shown in  \cite{Molchanov2014} and \cite{Baryshevsky2015}, respectively, refer to the instantaneous (peak) power rather than 
the average power we are using in the present article).

%\begin{figure}[tbh]
%	\centerline{{
%			\includegraphics[width=3 cm]{resonator_drawing.eps}}
%		\hfil
%		\includegraphics[width=3 cm]{resonator_photo.eps}%
%	} \caption{Resonator drawing (left) and photo (right)}
%	\label{fig:resonator}
%\end{figure}

%----------------
Experiments with the designed axial vircator\cite{Baryshevsky2015}
were focused on investigation of output power and operation stability for cathodes of different type 
and size, varied cathode-anode gap.

\begin{figure}[ht]
	\leavevmode
	\centering
	\resizebox{120mm}{!}{\includegraphics{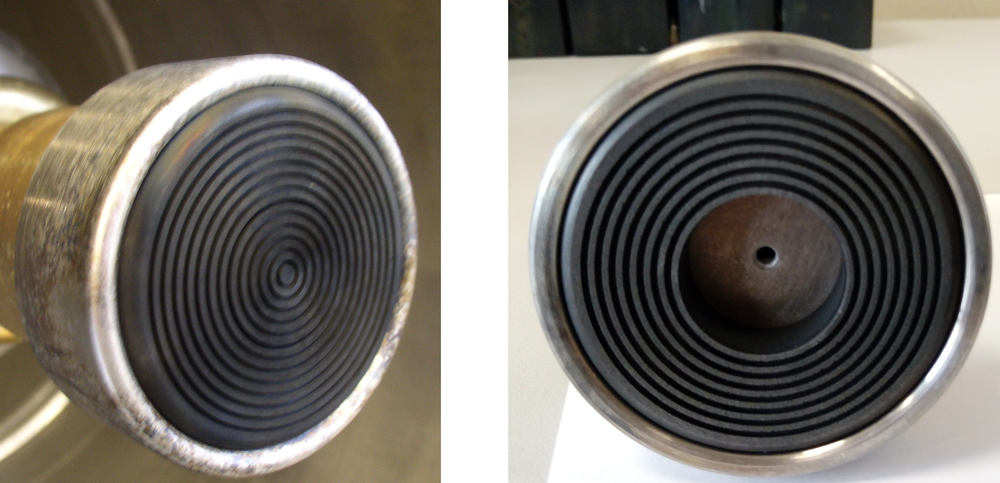}}\\
	\resizebox{120mm}{!}{\includegraphics{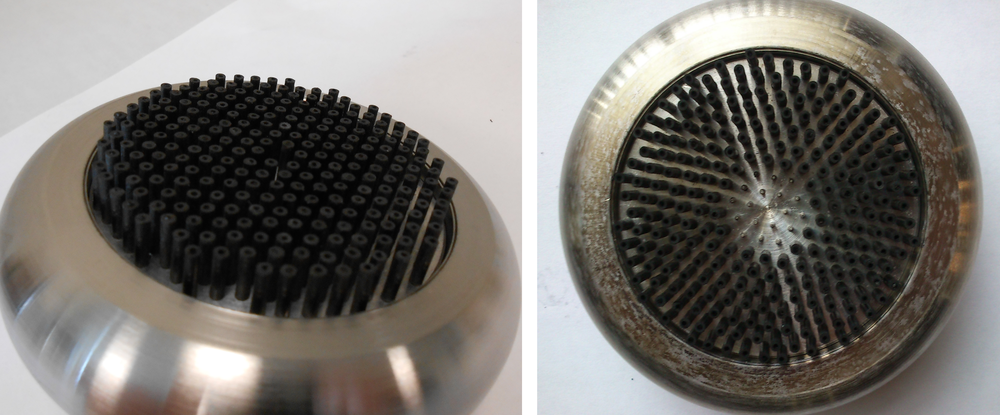}}\\
	\caption{Solid (top left) and ring-type (top right) cathodes made of dense fine grained graphite;
	solid (bottom left) and ring type (bottom right) multi-capillary cathodes}
	\label{fig:cathodes_axial}
\end{figure}

Solid and ring-type cathodes made of dense fine grained graphite,
and multi-capillary cathodes,  which emitter was formed by an array of carbon-epoxy capillaries 
were used (see Fig.~\ref{fig:cathodes_axial}). 
Outer diameter of graphite cathodes varied from 
60 to 71~mm;  inner diameter for ring-type graphite cathodes used in the experiments ranged from 20 to 26~mm. 
Engraved concentric grooves on flat emitting surface of each cathode facilitated uniform emission.
Two types of multicapillary cathodes with outer diameter of emitting area varying from $\sim 60$~mm to $\sim 70$~mm were tested:
the first one was similar to solid graphite cathodes (see Fig.~\ref{fig:cathodes_axial}, bottom left),
and the second one was similar to ring-type graphite cathodes (inner diameter of emitting area
was varied from  $\sim$21~mm to $\sim 28$~mm, see Fig.~\ref{fig:cathodes_axial}, bottom right).

%%TODO !!! CHECK
In all experiments with the solid graphite cathodes the observed power was lower than predicted in 
simulation~\cite{Molchanov2014}: $\sim 30$~MW for $\diameter$~60~mm cathode and 
 $\sim 60$~MW for cathode with $\diameter$~71~mm.

%В экспериментах со сплошными графитовыми катодами диаметром $60$~мм генерация была крайне нестабильной,
%а мощность излучения была намного ниже ($\sim 30$~MW), чем в моделировании~\cite{Molchanov2014}.
%Increase of solid graphite cathode diameter from 60~mm to 
%71~mm gave $\sim50$\% gain in microwave signal amplitude, что, однако, также
%существенно ниже результатов, предсказываемых\cite{Molchanov2014}.

\begin{figure}[th]
	\centering
	\resizebox{130mm}{!}{\includegraphics{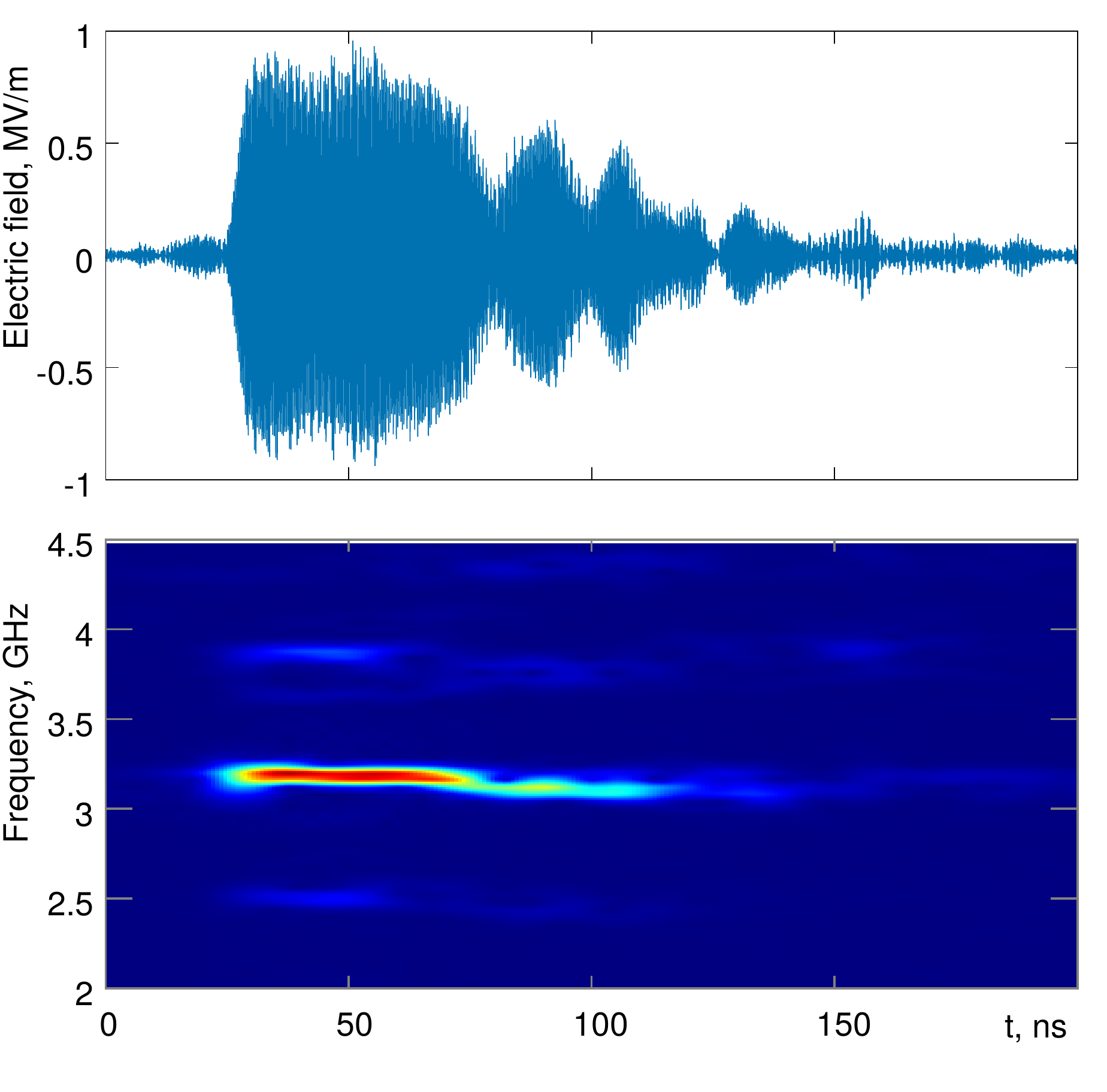}}\\
	\caption{Detected microwave signal with the highest amplitude and its spectrogram for the axial vircator with multicavity resonator,  graphite ring-type cathode with outer/inner diameters 60/24~mm and cathode-anode gap 16~mm \cite{Baryshevsky2015}}\label{fig:axial_signal}
\end{figure}

Ring-type graphite cathode with outer diameter 60 mm enabled us to achieve the power values
close to those predicted in \cite{Molchanov2014} (up to $300$~MW). 
The progressive enlargement of the cathode inner diameter from 20~mm to 24~mm, 
when keeping all other parameters the same, resulted in an increase of the radiation power to 
$150$~MW and even to $300$~MW in several shots (see Fig.~\ref{fig:axial_signal}). 
Increase of inner diameter from 24 to 26~mm led to decrease of microwave signal.
%%TODO translate
%
%
Duration of radiation pulse for shots with high output power attained 50~ns. Direct correlation of radiation pulse duration  with that for  current pulse was not observed: FWHM for current pulse was as high as 150--200 ns.
We explained this fact by high-voltage breakdown and plasma formation inside the resonator. 
%Длительность импульса в режиме максимальной мощности не превышала 50~нс.
%Следует обратить внимание, что длительность излучения, генерируемого аксиальным 
%виркатором, не зависела от длительности токового импульса, несмотря на то, 
%что его ширина на полувысоте варьировалась в диапазоне 150--200 нс. 
%Отсутствие 
%зависимости длительности микроволнового импульса от длительности токового импульса 
%свидетельствует о влиянии пробоев или перемыкания какод-анодного зазора на работу генератора.

Comparison of simulation results\cite{Molchanov2014} and 
data obtained experimentally\cite{Baryshevsky2015}
demonstrated perfect fit in basic frequency and accordance in radiated power.
However, the experiments demonstrated low reproducibility of results
though the operation parameters in the utmost and ordinary cases seemed to be identical. 
The subsequent analysis \cite{Gurnevich2015} explained the unstable operation by high 
dependence of axial vircator efficiency on the spread of the electron velocities in the beam:
growth of the momentum (energy) spread from 1\% to 5\% caused drop of the efficiency from 6\% to $\sim$1\%.

Multi-capillary cathodes were put for tests in anticipation of higher output due to 
more uniform electron beam with lower energy spread as compared to graphite cathodes.
However, radiation output produced by all variants of multi-capillary cathodes was the 
same as recorded in the shots with the solid cathodes made of dense fine grained graphite.

Experiments with the ring-type cathodes revealed one more challenge: we observed extremely fast 
damage of the anode mesh. Additional studies showed that this phenomenon was due to electrostatic 
cumulation of high-current electron beam~\cite{ElectrostaticCumulation2016,ElectrostaticCumulation2019}.

\FloatBarrier

%\subsection{Cathode-anode gap variation}

\begin{figure}[th]
	\centering
	\resizebox{140mm}{!}{\includegraphics{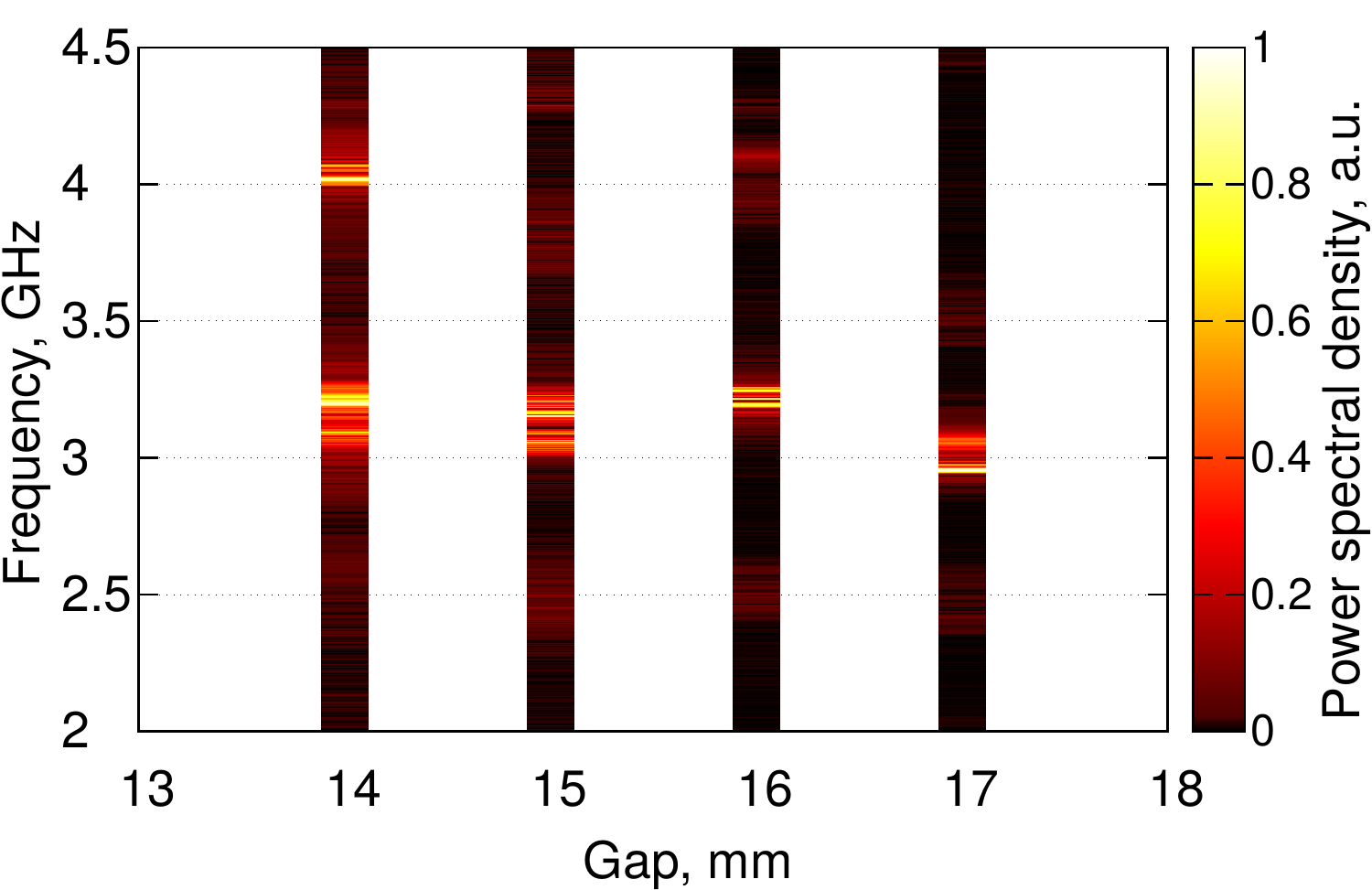}}\\
	\caption{Comparison of axial vircator spectra obtained for different cathode-anode gap values\cite{Baryshevsky2015} for solid graphite cathode of 60~mm diameter}\label{fig:specs_axial}
\end{figure}

According to simulation the cathode-anode gap mainly influences on the radiation spectra \cite{Anishchenko2014,Anishchenko2014_2}:
when the cathode-anode gap was reduced to appear below a certain value, the basic vircator frequency grew.  
%
%Отметим, что рост частоты не был плавным из-за наличия резонатора
%(собственные частоты которого образуют дискретный ряд). 
This is due to increase of electron-beam
plasma frequency \cite{Didenko1979,Sullivan1983}. 
We also observed such frequency behavior in the experiments\cite{Baryshevsky2015} (see Fig.~\ref{fig:specs_axial}).
%The experiments with cathode-anode gap variation were conducted for solid graphite cathode of 60~mm diameter.
%
With the cathode-anode gap varying from 14~mm to 17~mm,
the microwave pulse duration, and the peak power remained almost the same; but further increase
of gap led to power reduction and the microwave pulse delay
with respect to the current pulse. 
%%TODO translate
%
At the same time for each cathode-anode gap value the radiation frequency weakly depended on beam current.
%При этом при каждом фиксированном значении катод-анодного зазора 
%изменение тока пучка очень слабо влияло на частоту генерации.
%
%(изменению тока в пределах $\pm 10$\% соответствовало изменение частоты в пределеах не более чем $\pm 2$\%).

\FloatBarrier

%todo !!! CHECK

Stability of axial vircator operation
in our experiments was low and in many cases the radiated power was much lower than simulation had predicted.
The subsequent analysis \cite{Gurnevich2015} explained the
unstable operation by high dependence of source efficiency on the spread of electron velocities in the beam.
Nevertheless, by testing different cathode shapes, materials and operation conditions we succeeded  
to achieve efficiency  about 5\% and power up to 300~MW at 3.2~GHz. Ring-type graphite cathode with outer/inner diameters 60/24~mm produced radiation power from
$150$~MW to even $300$~MW in several shots. 
Duration of radiation pulse was about 50~ns at FWHM for current pulse as high as 150--200 ns.
We explained this fact by high-voltage breakdown and plasma expansion inside the resonator and cathode-anode gap. 
Extremely fast damage of the anode mesh was explained by electrostatic 
cumulation of high-current electron beam.
%%TODO translate
%Проанализировав эксперименты с аксиальным виркатором, можно заключить, 
%что импульсы микроволнового излучения не обладают стабильными мощностными 
%характеристиками, несмотря на то, что в отдельных выстрелах мощность излучения достигала 300~МВт. 
%Длительность микроволновых импульсов $\sim 50$~нс не зависела от длительности токовых импульсов, 
%менявшихся в диапазоне 150--200~нс, что свидетельствует о наличии преждевременных пробоев или 
%существенного влияния перемыкания катод-анодного зазора. 

\section{Reflex triode}
\label{sec:reflex}

%%todo from reviewer: 
%\textcolor{blue}{The virtual cathode oscillator in reflex triode geometry [5] demonstrated higher operating stability at high-power operation with longer pulse duration as compared with other sources. It provided good single frequency operation during the 100 ns pulse [6] (see Fig. 1). High operation stability was ensured by use of multi-capillary cathode [7]. Sealing of the HPM source was simulated only, since the vacuum pump drivers appeared to have been damaged by radiation during the tests.}
%%todo end from reviewer

A virtual cathode oscillator operating in reflex triode geometry~\cite{Didenko1979,Mankovski2004,Jiang2004,Liu2007,Grigoryiev2006,Zherlitsyn2007,Chen2007} 
has all the strengths and weaknesses typical for this type of HPM sources, particularly it demonstrates relatively low efficiency. 
However, introducing a resonant feedback into the system, one could enhance the efficiency (up to 10\%-12\%) and single-frequency operation \cite{Mankovski2004,Benford1987,Benford2007,Liu2008,Yang2010,ZhiQiang2008,Baryshevsky2017}. 
A vircator placed into a resonator, whose natural frequency is close to the virtual cathode oscillator frequency, attains the above resonant feedback. 
In \cite{Mankovski2004} the performance of a reflex triode was modified with the inclusion of reflecting strips  that
provided increase of  microwave peak power output to 330 MW at 11\% efficiency.

\begin{figure}[ht]
	\leavevmode
	\centering
	\includegraphics[width=100mm]{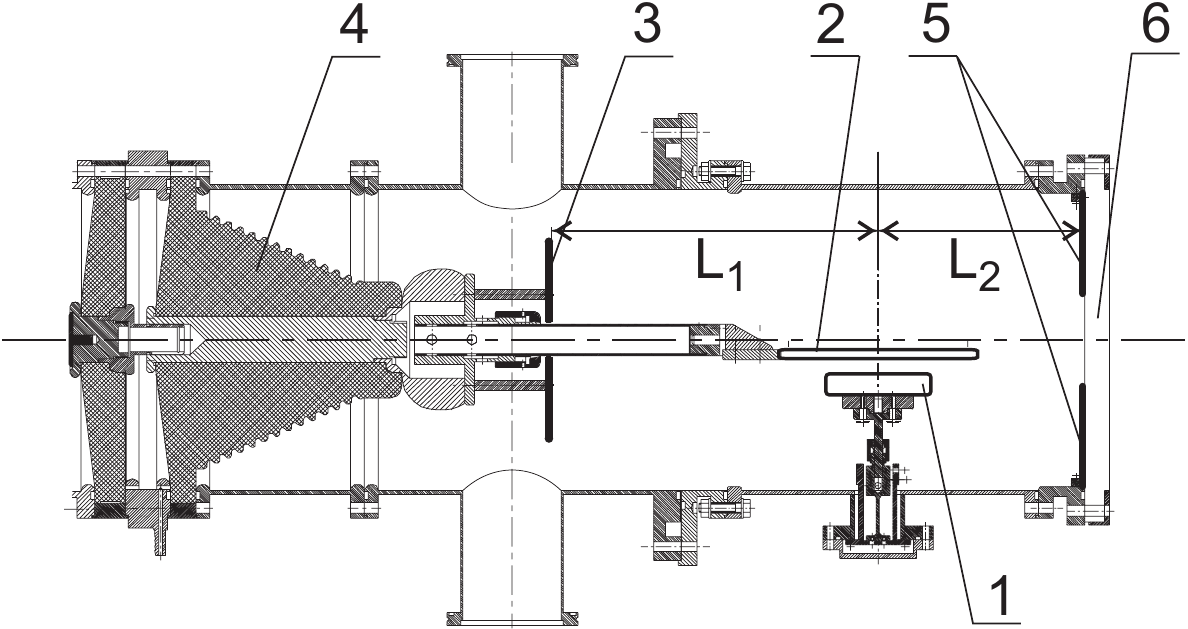}
	\includegraphics[width=90mm]{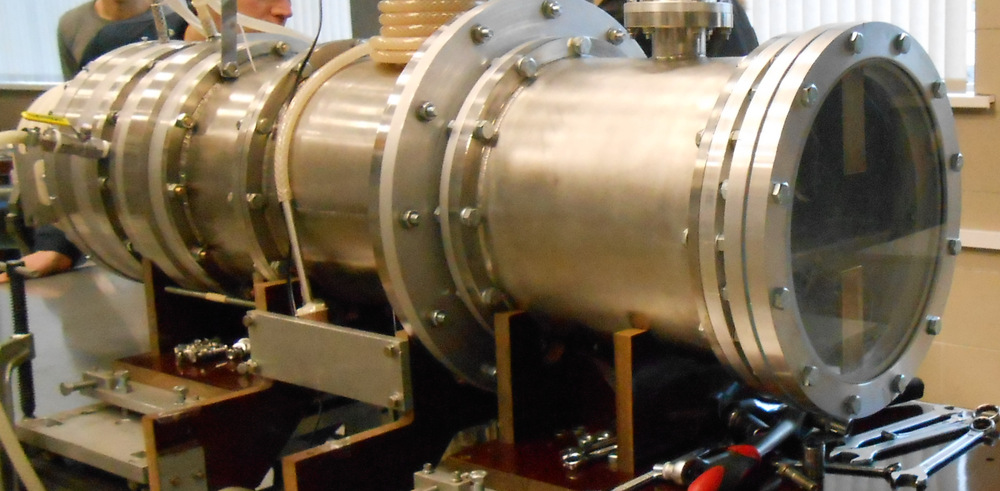}
	\caption{Reflex-triode geometry (top) and photo (bottom): 1 -- the explosive-emission cathode, 2 -- anode mesh and anode holder centered in the vacuum chamber, 3 -- disk-shaped reflector, 4 -- high-voltage vacuum feed-through,
		5 -- shiftable reflector with rectangular brass stripes, 6 -- output window}
	\label{fig:rt}
\end{figure}

Following the concept of vircator with resonant feedback
we developed the reflex triode with shiftable reflectors (see Fig.~\ref{fig:rt})  and experimentally studied its performance 
\cite{Baryshevsky2017}. 
Radiation frequency, output power and duration of radiation pulse were analyzed for different cathode-anode gap values, cathode materials and varied reflector positions.
System parameters ensuring high operating stability at single frequency ranging from 3.0 to 4.2 GHz and high-power operation with long pulse duration were obtained.

The reflex-triode geometry can be described as follows.
The vacuum chamber of 300 mm diameter encloses the triode reflex geometry vircator (see Fig. \ref{fig:rt}): $L_1$ and $L_2$ are the variable distances from the cathode axis to the disk-shaped and output shiftable/removable reflectors, respectively. Stainless steel woven anode mesh with 77\% geometric transparency was used in the experiments; the diameter of the mesh thread is 224 $\mu$m. 
Solid type cathodes were made of different materials: fine-grained graphite MPG-8, duralumin, steel nails, carbon-epoxy capillaries. 
The output reflector consisted of two rectangular brass stripes 40 mm in width and 100 mm in length, housed at a variable distance from the cathode axis, 
normally to the anode plane position.

The value of the cathode-anode gap was assured with 0.1~mm accuracy. Coplanarity of cathode and anode surfaces was controlled. The cathode-anode gap value and positions of both reflectors  (defined by $L_1$ and $L_2$ ) were tuned in order to observe stable single frequency generation and the highest output power.
A pulsed power supply, similar to that in \cite{Baryshevsky2015}, was used to drive the HPM source.

\begin{figure}[ht]
	\leavevmode
	\centering
	\resizebox{130mm}{!}{\includegraphics{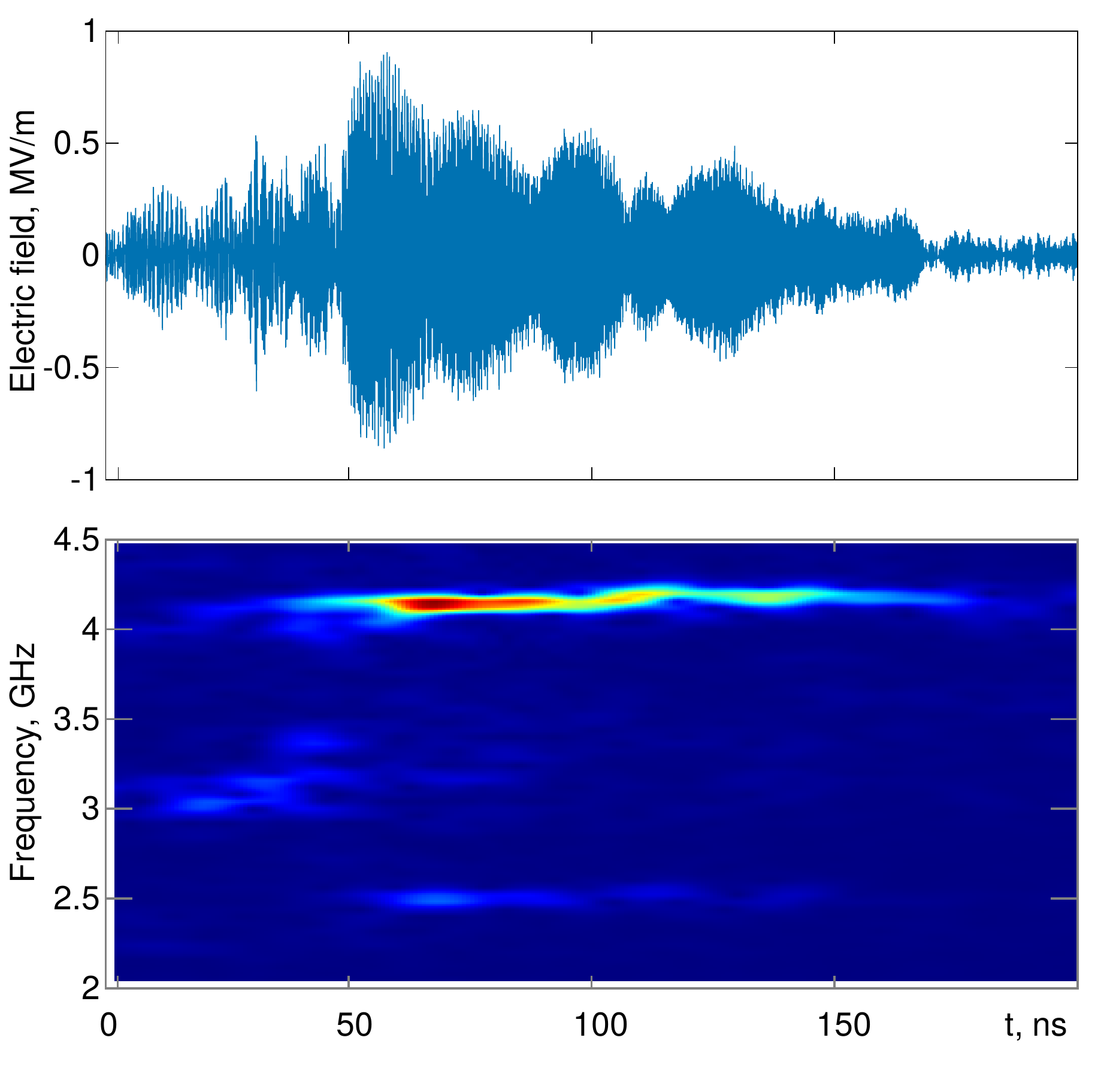}}\\
	\caption{Detected microwave signal with the highest amplitude and its spectrogram}
	\label{fig:rtmax}
\end{figure}

\begin{figure}[ht]
	\leavevmode
	\centering
	\resizebox{100mm}{!}{\includegraphics{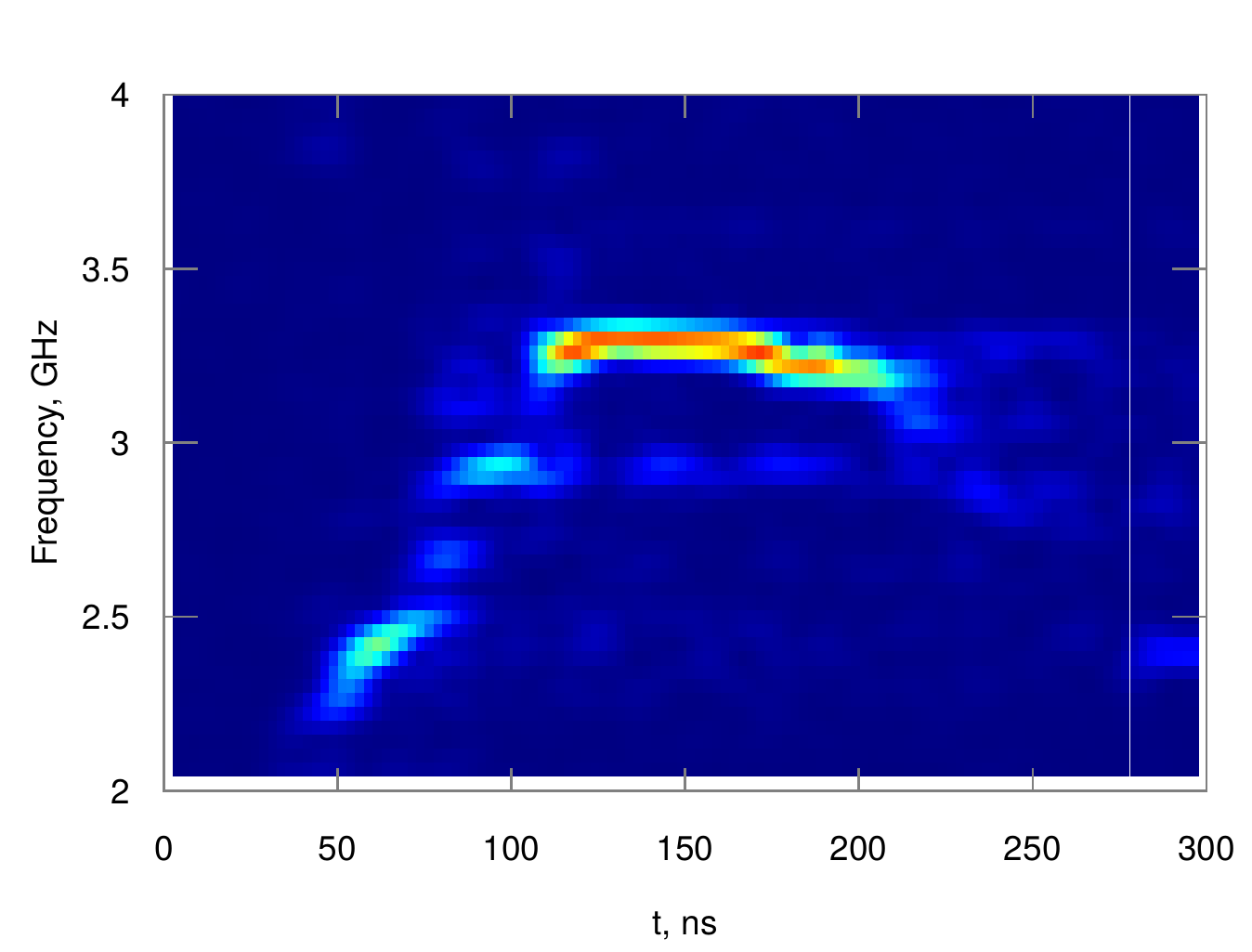}}\\
	\caption{Spectrogram of the detected microwave signal at the most stable operation regime}
	\label{fig:rtstable}
\end{figure}

\begin{figure}[ht]
	\leavevmode
	\centering
	\resizebox{80mm}{!}{\includegraphics{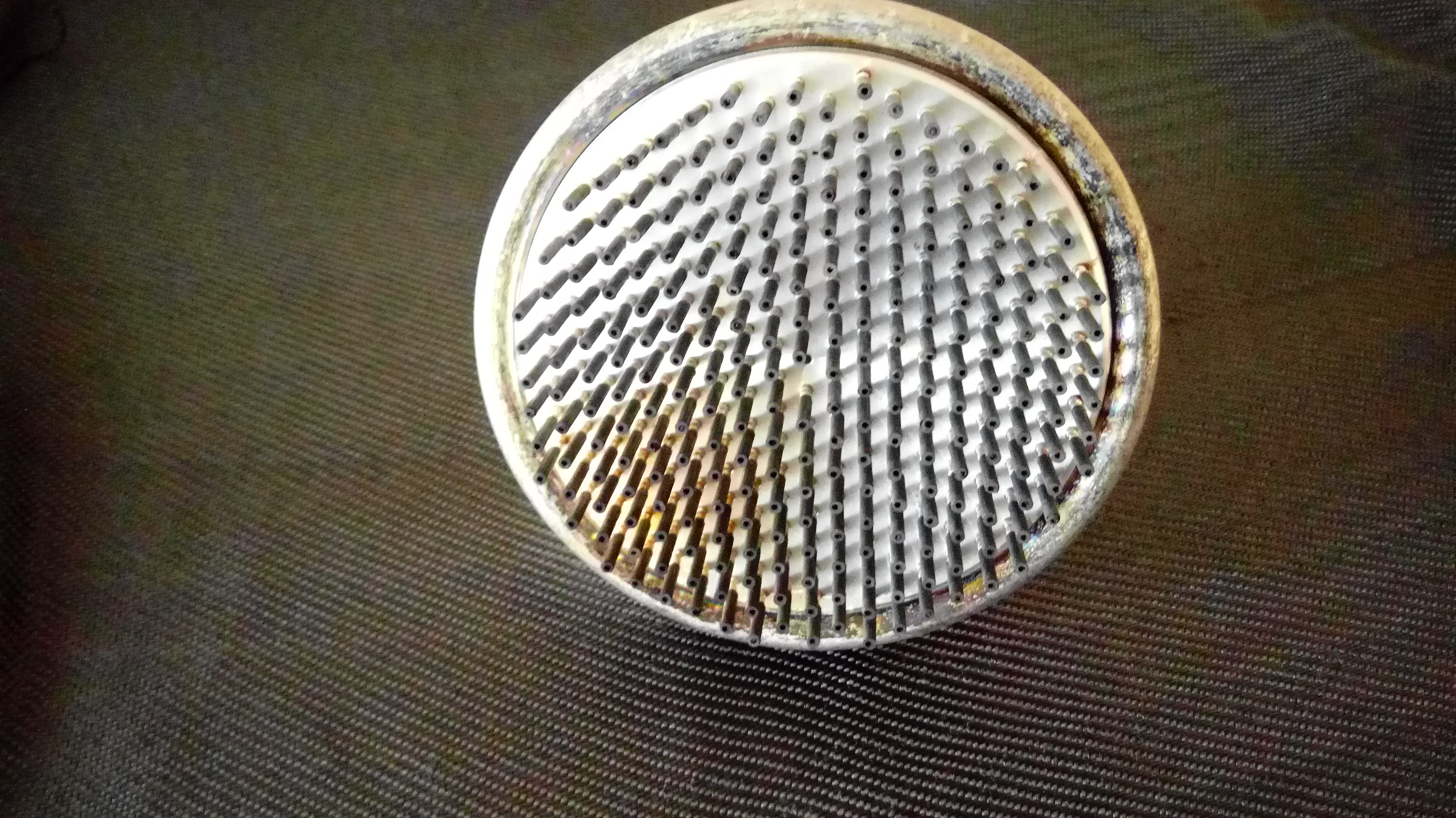}}\\
	\caption{Multicapillary cathode}
	\label{fig:cathode}
\end{figure}

The maximum radiated power was observed at cathode-anode gap 16~mm and reflector positions $L_1=290$ mm and $L_2=164$~mm.
 As much as 460~MW
was produced at 4.16~GHz single-frequency operation at maximum diode voltage 460~kV and amplitude of electron beam current 18~kA. 
%%TODO translate
%This value is close those observed in experiments with axial vircator.
% 
%По своей величине она близка к амплитуде тока в диоде аксиального виркатора в режиме максимальной мощности (20~кА).
%
 Detected microwave signal with electric field strength amplitude $\sim800$ kV/m @ 1 m and its spectrogram are shown in Fig. \ref{fig:rtmax}. 
 The cathode-anode gap increase to 17 and 18 mm resulted in spectrum broadening and shift of its central frequency.

The most stable operation of the reflex triode was observed when output reflector was removed. The corresponding spectrogram is shown in Fig. \ref{fig:rtstable}. The main feature of the configuration is the usage of multicapillary cathodes (see Fig. \ref{fig:cathode}) described in \cite{Gleizer2013}. It is found that this type of cathodes provides the most  stable  single-frequency operation at 200~MW.

%TODO !!! CHECK

Duration of radiation pulse correlated with FWHM of current pulse: when duration of current pulse was increased from 150 to 250~ns, the microwave pulse duration grew from 50 to 150~ns.  
This was in contrast to operation of axial vircator,  where high-voltage breakdowns and plasma expansion inside the resonator and cathode-anode gap limited duration of radiation pulse.

%Следует обратить внимание, что длительность излучения, генерируемого отражательным триодом, зависела от длительности токового импульса. Его ширина на полувысоте варьировалась в диапазоне 150--250 нс, что приводила к изменению длительности микроволнового импульса от 50 до 150 нс без снижения мощности. Линейная зависимость длительности микроволнового импульса от токового импульса свидетельствует об отсутствии влияния пробоев и замыкания зазора на работу генератора.
%

It should be noted that in order to avoid the impact of powerful microwave radiation on the vacuum pumps, they are turned off before every shot. At this time, evacuation is carried out by getter pumps. Moreover, the presence of vacuum pumps make us try-out sealing of the reflex triode in its design.

%\subsection{Cathode-anode gap}

\begin{figure}[th]
	\centering
	\resizebox{140mm}{!}{\includegraphics{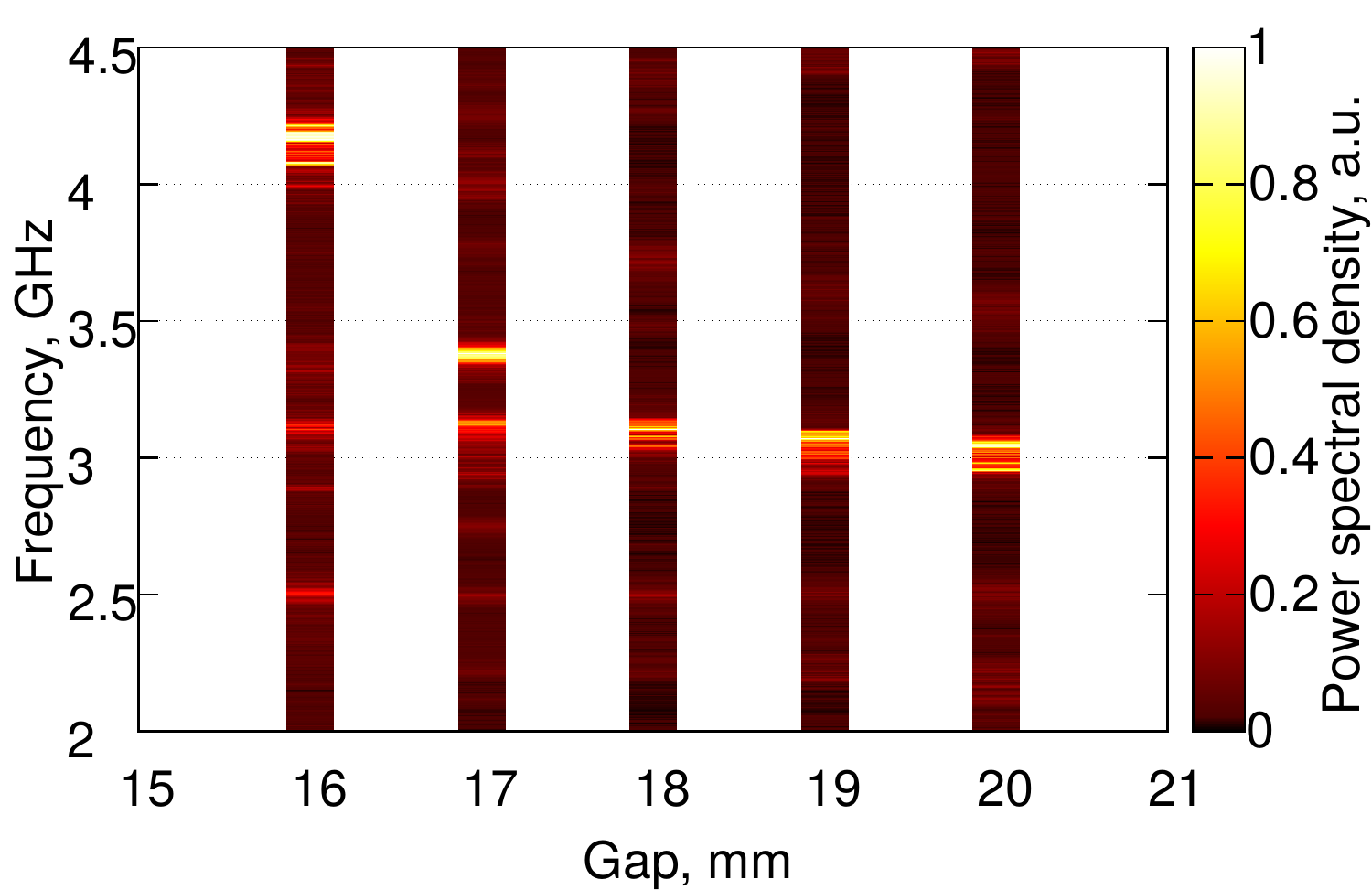}}\\
	\caption{Comparison of reflex triode spectra obtained for different cathode-anode gap values.}\label{fig:9}
\end{figure}

Variation of the cathode-anode gap value over the range
20--16 mm demonstrated change of the main radiation
frequency within the range 3.0--4.16 GHz (see Fig.~\ref{fig:9}).
In each diagram zone, which corresponds to the fixed cathode-anode
gap value, the sum of frequency spectra obtained in
several experiments under similar conditions is presented.
%TODO !!! CHECK
%
In all the experiments the frequency of radiation  was mainly determined by the cathode-anode gap  $d$, rather than beam current. Frequency was proportional to $\sim 1/d$ with small variations due to current change, for example: for cathode-anode gap 16~mm change of the beam current from 15 to 18.5~kA 
(i.e. $16.8$~kA$\pm 10$\%) resulted in frequency variations within $\pm 2$\% range (see Fig.~\ref{fig:9}).
%
%Отметим, что во всех эксприментах частота генерации определялась
%в первую очередь именно величиной катод-анодного зазора $d$ (она была приблизительно пропорционально $\sim 1/d$),
%а не величиной тока пучка.
%Изменение же параметров питающего импульса, например, амплитуды тока в диоде
%при фиксированном значении зазора приводило лишь к небольшим вариациям частоты;
%например, при катод-анодном зазоре 16~мм изменению тока в диапазоне от 15 до 18.5~kA 
%(т.е. $16.8$~kA$\pm 10$\%) соответствовали колебания частоты в 
%пределах не более чем $\pm 2$\%, что хорошо видно на рис.~\ref{fig:9}).

%
% The
%change of the cathode anode gap value from 17 to 16 mm at
%a fixed position of the output reflector leads to the hop of
%the radiation frequency from 3.37 to 4.16 GHz, which is also
%shown in Fig.~\ref{fig:9}.

\FloatBarrier

%TODO !!! CHECK
Thus, HPM pulses produced by the reflex triode with the multicapillary cathode had stable frequency and power about 200~MW.
Duration of current pulse determined that of radiation pulse: increase of current FWHM from 150 to 250~ns caused growth of microwave pulse duration  from 50 to 150~ns.

%Таким образом, можно заключить, что импульсы микроволнового излучения, генерируемые 
%отражательным триодом с мультикапиллярным катодом, обладают стабильными мощностными 
%и частотными характеристиками. Мощность излучения составляет 200 МВт. Длительность 
%этих импульсов, лежащая в диапазоне 50--150~нс, линейно связана с длительностью токовых 
%импульсов, изменяющихся от 150 до 250~нс, что свидетельствует об отсутствии влияния 
%пробоев и перемыкания катод-анодного зазора на работу генератора. 

\section{Conclusion}

We compared three types of compact HPM sources capable to produce high output power, long pulse duration and good operating stability
operating without a magnetic field (see Table~\ref{tab:comparison}).

\begin{table}[h!]
	\begin{center}
		\begin{tabular}{|c|c|c|c|}
			\hline
			Parameter & Axial vircator & SCO & Reflex triode \\
			\hline
			Current pulse FWHM [ns] & 150--200 & 200 & 150--250\\
			Radiation power [MW] & 300 & 30 & 200 (max.\,460)\\
			Duration of microwave pulse [ns] & 50 & 5-10 & 50--150\\
			Operation stability & low & high & high\\
			\hline
		\end{tabular}
		\caption{Comparison of different HPM sources.}
		\label{tab:comparison}
	\end{center}
\end{table}

For split-cavity oscillator we observed high stability, but low radiated power and pulse duration. Frequency was determined by fundamental mode in SCO cavity.
Pulse duration was limited by breakdowns inside the cavity to be as low as 10~ns.

For axial 
vircator experiments demonstrated 
low reproducibility with infrequent shots with high power and efficiency: the radiated power about 300~MW was observed 
in several ``best'' shots, which were less than 10\% in the total amount of all tests carried in the equivalent conditions.
%, the rest of more than 90\% gave twice as lower result (300 MW).
%
Radiation spectra was determined by cathode-anode gap:
when the cathode-anode gap was reduced to appear below a certain value, the basic vircator frequency grew.
Experimentally observed spectra fitted well the simulation results \cite{Molchanov2014}.
Duration of radiation pulse for shots with high output power attained 50~ns. Direct correlation of radiation pulse duration  with that for  current pulse was not observed.

Thus, HPM pulses produced by the reflex triode with the multicapillary cathode had stable frequency and power about 200~MW. As much as 460~MW
was produced at 4.16~GHz single-frequency operation at maximum diode voltage 460~kV and amplitude of electron beam current 18~kA. 
Duration of current pulse determined that of radiation pulse: increase of current FWHM from 150 to 250~ns caused growth of microwave pulse duration  from 50 to 150~ns.
Radiation frequency was proportional to $\sim 1/d$ with small variations due to current change, namely: $\pm 10$\% change in current resulted in frequency variations within $\pm 2$\% range.

According to the above analysis reflex triode bears the palm. Nevertheless, two other competitors still have good opportunities in case of proper considering their weak points.


\begin{thebibliography}{2008}
%\bibliographystyle{PhdThesis}

%Introduction
\bibitem{Marder1992}
B.M. Marder, "The split-cavity oscillator: a high-power e-
beam modulator and microwave source," IEEE Trans. Plasma
Sci., vol. 20 (1992) no.3, pp.~312--331.

\bibitem{Miller1982}
R. B. Miller, An Introduction to the Physics of Intense Charged Particle
Beams. New York, NY, USA: Springer-Verlag, 1982.

\bibitem{Granatstein1987}
V. L. Granatstein and I. Alexeff, Eds., High-Power Microwave Sources.
Boston, MA, USA: Artech House, 1987.

\bibitem{Benford2007}
J. Benford, J. A. Swegle, and E. Schamiloglu, High Power Microwaves,
2nd ed. Boca Raton, FL, USA: CRC Press, 2007.

\bibitem{Didenko1979}
A. N. Didenko et al., "Generation of high power rf pulses in the
magnetron and reflex triode systems," in Proc. 3rd Int. Topical Conf.
High-Power Electron Ion Beam Res. Technol., vol. 2. Novosibirsk,
Russia (1979) pp. 683--691.

\bibitem{Sullivan1983}
D. J. Sullivan, "High power microwave generation from a virtual
cathode oscillator (vircator)", IEEE Trans. Nucl. Sci., vol. 30 (1983) no.4, pp.~3426--3428.

%SCO


\bibitem{Lemke1994}
Raymond W. Lemke, M. Collins Clark, Barry M. Marder, ''Theoretical and experimental investigation of a method for increasing the output power of a microwave tube based on the split-cavity oscillator'', Journal of Applied Physics.  V75, (1994), no.10, pp.~5423--5432.

\bibitem{Fan2004} Fan Zhikai, Liu Qingxiang, Chen Daibing, Tan Jie, Zhou Haijing. ''Theoretical and experimental researches on C-band three-cavity transit-time effect oscillator'', Science in China Ser. G Physics, Mechanics and Astronomy.  V47 (2004) no.3. pp.~310-329.

\bibitem{He2004} He Jun-Tao, Zhong Hui-Huang, Qian Bao-Liang, Liu Yong-Gui. A new method for increasing output power of a three-cavity transit time oscillator. / Chinese Physics Letters. 2004. V21. №7. P. 1302–1305.

\bibitem{Bar14}	Relativistic Split-Cavity Oscillator/ V.G. Baryshevsky. – Research Institute for Nuclear Problems, 2014. – Mode of access: arxiv.org/abs/1402.3403 Date of access: 05.01.2016.

\bibitem{Moroz2018}
I. Moroz, A. Rouba, Theoretical and experimental research of split-cavity oscillator, Theoretical and experimental research of split-cavity oscillator : thesis of 61st International Conference for Students of Physics and Natural Sciences <<Open Readings 2018>> , Vilnius, March 20-23, 2018, p. 111.

\bibitem{Moroz2020}
S. Anishchenko, I. Moroz, A. Rouba, Radiation instability in relativistic split-cavity oscillator, Radiation instability in relativistic split-cavity oscillator: thesis of 63st International Conference for Students of Physics and Natural Sciences “Open Readings 2020” , Vilnius, March 17-20, 2020, p. 398.

\bibitem{Bogd2004}
I. L. Bogdankevich et al. Influence of the Electrons Reflected from the Collector on the Parameters of a High-Current Relativistic Electron Beam. Plasma Physics Reports, Vol. 30, No. 5, 2004, pp. 376–382.

\bibitem{Calico}
Calico S.E. et al. Experimental and theoretical investigations of a magnetically insulated line oscillator, SPIE Vol. 2557. P. 50

\bibitem{Haworth1998}
Haworth M.D. et al. Significant pulse-lengthening in a multigigawatt magnetically insulated transmission line oscillator, IEEE Trans. Plasma Sci. 1998. Vol. 26. No. 3. P. 312


%Vircators
\bibitem{Jiang2004}
W. Jiang, N. Shimada, S. D. Prasad, and K. Yatsui, "Experimental and
simulation studies of new configuration of virtual cathode oscillator,"
IEEE Trans. Plasma Sci., vol. 32, no. 1, pp. 54--59, Feb. 2004.

\bibitem{Kitsanov2002}
S. A. Kitsanov et al., "S-band vircator with electron beam premodulation
based on compact pulse driver with inductive energy storage," IEEE
Trans. Plasma Sci., vol. 30, no. 3, pp. 1179--1185, Jun. 2002.

\bibitem{Didenko1996}
A. N. Didenko et al., "Reflex triode with resonant cavity as a load for
inductive storage," in Proc. 11th Int. Conf. High-Power Particle Beams,
vol. 1. Jun. 1996, pp. 445--448.

\bibitem{Champeaux2013}
S. Champeaux, P. Gouard, R. Cousin, and J. Larour, "Numerical
evaluation of the role of reflectors to maximize the power efficiency
of an axial vircator," in Proc. 14th Int. Vac. Electron. Conf. (IVEC),
May 2013, pp. 1--2.

\bibitem{Zhenxiang2003}
L. Zhenxiang, S. Ting, Z. Jiande, and Q. Baoliang, "Particle simulation
of an improved axially extracted vircator," Plasma Sci. Technol., vol. 5,
no. 5, pp. 2007--2010, 2003.

\bibitem{Ting2002}
S. Ting, W. Yong, Q. Bao-Liang, and T. Qi-Mei, "A compact vircator
with feedback annulus operated in quasi-single TM 01 mode within the
C band," Chin. Phys. Lett., vol. 19, no. 11, pp. 1646--1649, 2002.

\bibitem{ZhiQiang2008}
L. Zhi-Qiang et al., "Simulation and experimental research of
a novel vircator," Chin. Phys. Lett., vol. 25, no. 7, pp. 2566--
2569, 2008.

%\bibitem{Baryshevsky2013}
%V. Baryshevsky, A. Gurinovich, P. Molchanov, S. Anishchenko, and
%E. Gurnevich, "2-D simulation and experimental investigation of an axial
%vircator," IEEE Trans. Plasma Sci., vol. 41, no. 10, pp. 2712--2716,
%Oct. 2013.

\bibitem{Baryshevsky2015}
V. Baryshevsky, A. Gurinovich, E. Gurnevich, P. Molchanov,
``Experimental study of an axial vircator with resonant cavity,''
IEEE Trans. Plasma Sci., vol. 43, no. 10, pp. 3507--3511, 2015.

\bibitem{Molchanov2014}
P. V. Molchanov, E. A. Gurnevich, V. V. Tikhomirov, and S. E. Siahlo.
(2014). "Simulation of an axial vircator with a three-cavity resonator."
[Online]. Available: http://arxiv.org/abs/1408.1824

\bibitem{Gurnevich2015}
E. Gurnevich, P. Molchanov, "The effect of the electron-beam parameter spread on microwave generation in a three-cavity axial vircator," IEEE Trans. Plasma Sci., vol. 43, no. 4, pp.
1014--1017, 2015.

\bibitem{ElectrostaticCumulation2016}
S. Anishchenko, V. Baryshevsky, N. Belous, A. Gurinovich, E. Gurnevich, P. Molchanov, "Cumulation of high-current electron beams: theory and experiment," vol. 45, no. 10, pp. 2739--2743, 2016.

\bibitem{ElectrostaticCumulation2019}
S.V. Anishchenko, V.G. Baryshevsky, A.A. Gurinovich, "Electrostatic cumulation of high-current electron beams for terahertz sources," vol. 22, 043403, 2019.


\bibitem{Anishchenko2014}
S. V. Anishchenko and A. A. Gurinovich, "Modeling of high-current
devices with explosive electron emission," Comput. Sci. Discovery,
vol. 7, no. 1, p. 015007, 2014.

\bibitem{Anishchenko2014_2}
S. V. Anishchenko and A. A. Gurinovich,
``Modeling of explosive electron emission and electron beam dynamics in high-current devices,''
J. Phys., Conf. Ser., vol. 490, no. 1, p. 012116, 2014.


%REFLEX
\bibitem{Mankovski2004}
J. J. Mankowski, X. Chen, J. C. Dickens, and M. Kristiansen, "Exper-
imental optimization of a reflex triode virtual cathode oscillator," in
Proc. Int. Conf. High-Power Particle Beams (BEAMS), Jul. 2004,
pp. 426–429.

\bibitem{Liu2007}
L. Liu, L. M. Li, X. P. Zhang, J. C. Wen, H. Wan, and Y. Z. Zhang,
"Efficiency enhancement of reflex triode virtual cathode oscillator using
the carbon fiber cathode," IEEE Trans. Plasma Sci., vol. 35, no. 2,
pp. 361–368, Apr. 2007.

\bibitem{Grigoryiev2006}
V. P. Grigoryiev, A. G. Zherlitsyn, T. V. Koval, G. V. Melnikov, and
P. Ya Isakov, "Mode structure research of a field in the triode with
the virtual cathode with an active feedback," in Proc. 14th Symp. High
Current Electron., Tomsk, Russia, 2006, vol. 40. no. 11, pp. 372–375.

%\bibitem{Zherlitsyn2007}
%A. G. Zherlitsyn, G. V. Mel’nikov, and P. Ya Isakov, "Effect of
%feedback on the microwave radiation in a triode with a virtual cathode,"
%J. Commun. Technol. Electron., vol. 52, no. 7, pp. 798–802, 2007.

\bibitem{Chen2007}
Y. Chen, J. Mankowski, J. Walter, M. Kristiansen, and R. Gale, "Cathode
and anode optimization in a virtual cathode oscillator," IEEE Trans.
Dielectr. Electr. Insul., vol. 14, no. 4, pp. 1037–1044, Aug. 2007.

\bibitem{Zherlitsyn2007}
A.G. Zherlitsyn, G.V. Mel’nikov, P.Ya. Isakov, "Effect of
feedback on the microwave radiation in a triode with a virtual
cathode," J. Commun. Technol. and Electron., vol. 52, no. 7, pp.
798--802, 2007.

\bibitem{Benford1987}
J. Benford, D. Price, H. Sze, and D. Bromley, "Interaction of a vircator
microwave generator with an enclosing resonant cavity," J. Appl. Phys.,
vol. 61, no. 5, pp. 2098–2100, 1987.

\bibitem{Liu2008}
G. Z. Liu et al., "Coaxial cavity vircator with enhanced efficiency,"
J. Plasma Phys., vol. 74, no. 2, pp. 233–244, 2008.

\bibitem{Yang2010}
W. Yang, Z. Dong, and Y. Dong, "Numerical studies of a new-type
vircator with a resonant cavity," IEEE Trans. Plasma Sci., vol. 38, no. 9,
pp. 2428–2433, Sep. 2010.


\bibitem{Baryshevsky2017}
V. Baryshevsky, A. Gurinovich, E. Gurnevich, and P.
Molchanov, "Experimental study of a triode reflex geometry
vircator," IEEE Trans. Plasma Sci., vol. 45, pp. 631--635, 2017.

% \bibitem{Krasik1}
%J.~Gleizer, T.~Queller, Y.~Bliokh, S.~Yatom, V.~Vekselman, Y.~Krasik, and
%V.~Bernshtam, ``High-current carbon-epoxy capillary cathode,'' 
%\emph{J. Appl. Phys.}, {\bf112}(2), p.~023303, 2012.

\bibitem{Gleizer2013}
T.~Queller, J.~Gleizer, and Y.~Krasik, ``High-current long-duration uniform
electron beam generation in a diode with multicapillary carbon-epoxy
cathode,'' \emph{J. Appl. Phys.}, vol. {\bf114}, 123303, 2013.

%\bibitem{Gurnevich2016} V.G.~Baryshevsky,  N.A.~Belous, A.M.~Belov, A.A.~Gurinovich, E.A.~Gurinovich,
%E.A.~Gurnevich, P.V.~Molchanov,
%``Generation of high-current electron beams in diodes with large-area explosive emission cathodes,''
%\emph{IEEE Transactions on Plasma Science}, {\bf44}(7), pp.~1103--1111, 2016.
%
%\bibitem{Miller1998}
%R.~B. Miller, ``Mechanism of explosive electron emission for dielectric fiber
%(velvet) cathodes,'' \emph{J. Appl. Phys.}, {\bf84}(7),  p.~3880, 1998.
%
%
%\bibitem{Krasik2001}
%Y.~E. Krasik, A.~Dunaevsky, A.~Krokhmal, J.~Felsteiner, A.~V. Gunin, I.~V.
%Pegel, and S.~D. Korovin, ``Emission properties of different cathodes at
%$e\leqslant 10^5$~v/cm,'' \emph{J. Appl. Phys.}, {\bf89}(4), pp.~2379--2399, 2001.
%
%\bibitem{Shiffler2008}
%D.~Shiffler, M.~Haworth, K.~Cartwright, R.~Umstattd, M.~Ruebush, S.~Heidger,
%M.~LaCour, K.~Golby, D.~Sullivan, P.~Duselis, and J.~Luginsland, ``Review of
%cold cathode research at the air force research laboratory,'' \emph{IEEE
%	Trans. Plasma Sci.}, {\bf36}(3), pp.~718--728, 2008.
%
%
%\bibitem{Roy2009}
%A.~Roy, R.~Menon, S.~Mitra, S.~Kumar, V.~Sharma, K.~V. Nagesh, K.~C. Mittal,
%and D.~P. Chakravarthy, ``Plasma expansion and fast gap closure in a high
%power electron beam diode,'' \emph{Phys. Plasmas}, {\bf16}(5), p.~053103, 2009. 

\end{thebibliography}
\end{document}